\documentclass[aps,prl,twocolumn,showpacs,preprintnumbers,amsmath,amssymb]{revtex4-1}
\usepackage{ifpdf}
 \newif\ifpdf
\ifx\pdfoutput\undefined
   \pdffalse
\else
   \pdfoutput=1
   \pdftrue
\fi
\ifpdf
   \usepackage{graphicx}
   \usepackage{epstopdf}
   \usepackage{color}
\else
   \usepackage{graphicx}
\fi
\usepackage{bm}
\usepackage{srctex}
\usepackage{hyperref}
\DeclareMathOperator{\Ss}{S_{ \scriptscriptstyle\Sigma}}

\DeclareMathOperator{\HKH}{\mathit{H}_{\mathrm{\scriptscriptstyle{KK}}}}
\DeclareMathOperator{\EFM}{\mathit{E}_{\mathrm{\scriptscriptstyle{FM}}}}
\DeclareMathOperator{\EAFM}{\mathit{E}_{\mathrm{\scriptscriptstyle{AFM}}}}
\DeclareMathOperator{\ENEM}{\mathit{E}_{\mathrm{\scriptscriptstyle{NEM}}}}
\DeclareMathOperator{\EDIM}{\mathit{E}_{\mathrm{\scriptscriptstyle{DIM}}}}
\begin{document}
\newcommand{\remark}[1] {\noindent\framebox{
\begin{minipage}{0.96\columnwidth}\textbf{\textit{ #1}}
\end{minipage}}
}

\newcommand{\question}[1] {\noindent\framebox{
\begin{minipage}{0.96\columnwidth}\textbf{\textit{Q: #1}}
\end{minipage}}
}

\newcommand{\bnabla}{{\boldsymbol{\nabla}}}
\newcommand{\eff}{\mathrm{eff}}

\title{ Interplay of localisation and competing interaction channels: cascade of quantum phase transitions
}

\author{A.~M.~Belemuk}
\affiliation{Institute for High Pressure Physics, Russian Academy of Science, Troitsk 142190, Russia}
\affiliation{Department of Theoretical Physics, Moscow Institute of Physics and Technology, 141700 Moscow, Russia}

\author{N.~M.~Chtchelkatchev}
\affiliation{L.D. Landau Institute for Theoretical Physics, Russian Academy of Sciences,117940 Moscow, Russia}
\affiliation{Department of Theoretical Physics, Moscow Institute of Physics and Technology, Moscow 141700, Russia}
\affiliation{Institute for High Pressure Physics, Russian Academy of Science, Troitsk 142190, Russia}

\author{A.~V.~Mikheyenkov}
\affiliation{Institute for High Pressure Physics, Russian Academy of Science, Troitsk 142190, Russia}
\affiliation{Department of Theoretical Physics, Moscow Institute of Physics and Technology, 141700 Moscow, Russia}

\author{Wu-Ming Liu}
\affiliation{Beijing National Laboratory for Condensed Matter Physics, Institute of Physics, Chinese Academy of Sciences, Beijing 100190, China}

\date{\today}
\begin{abstract}
We investigate the interplay of localization, interactions and (pseudo)spin degrees of freedom on quantum states of particles on the lattice. Our results show that breaking the paradigm density-density interaction $U_0\gg$ (pseudo)spin-(pseudo)spin interaction $U_s$ will drive the sequence of quantum phase transitions (QPT), where (pseudo)spin state and particle ordering, in case of several particle species, on the lattice are strongly changed. QPT driven by competing interactions, $|U_s|\sim U_0$, manifest itself in singularities of effective exchange integrals. $|U_s|\sim U_0$ implies a frustration  when the interactions standing alone drive the system to different phases. Even at $U_s=0$, there is typically a QPT induced by $U_s$ sign change. Vector cold atoms, Fermions or Bosons, on optical lattices are the state-of-the-art realization of our system where $U_s$ is tunable \textit{in situ}.
\end{abstract}


\maketitle

Ultracold atoms in optical lattices have opened fascinating possibilities for the controlled investigation and simulation of strongly correlated systems~\cite{Kuklov2003PRL,George14_RoMP,dutta2014RMP,Thywissen2016PhysRevA}. Unlike solid state systems, unique flexibility of the optical lattices enables continuous tuning of interaction parameters driving quantum atom states between the localised and delocalised phases and within them~\cite{Greiner2002Nature,Kohl2005PRL,kawaguchi2012spinor,hauke2012}. A particularly challenging and interesting problem  concerns the system with a number of  competing interaction channels. This leads to frustration and induces the  sequence of quantum phase transitions with  abundant phase diagram.

Here we focus on strongly correlated particles, Bosons or Fermions, on the lattice where particle density, spin and/or another degree of freedom (pseudospin) generate the competing interaction channels~\cite{anderlini2007Nature,folling2007Nature,anderlini2007Nature,trotzky2008Science,Wagner2012PRA,Chtchelkatchev2014PRA}. The kinetic energy (intersite hopping) in our investigation is much smaller than interactions. In this case typically energy gap opens and particles become  localised in the Mott insulating phase~\cite{gebhard2003mott,lewenstein2012book,George14_RoMP}. Mott insulators are of growing interest in advanced physics research, and are not yet fully understood~\cite{Lewenstein2015RepProgPhys}. The interplay of localisation and competition of interactions  generate new phases on top of (pseudo)spin degrees of freedom.

Low energy properties of bosonic (or fermionic) atoms in a optical lattice are well described by the Hubbard-like Hamiltonian that basically consists of the kinetic-part characterized by the hopping amplitudes $t$ and the single-site interaction~\cite{Bruder1998PRL,levin2012book}. For Bosonic atoms the interaction can be  decomposed into the density-density repulsion and (pseudo)spin-dependent part with interaction constants $U_0$ and $U_s$ respectively~\cite{Lewenstein2015RepProgPhys}.  For Fermionic atoms Hubbard $U_0$-repulsion is typically spin dependent~\cite{gebhard2003mott}: it is the product of particle densities corresponding to opposite spin, while $U_s$-part usually develops when there is a pseudospin degree of freedom in addition to spin. Pseudospin describes atom species when particle system includes several sorts of atoms~\cite{Kuklov2003PRL}. Other possibility includes ``orbital'' degrees of freedom of different origin~\cite{li2013Nature}.

So, we focus on the Mott  insulators parameter space, where $U_0\gg t$~\cite{kawaguchi2012spinor}, and show that breaking the usual paradigm $U_0\gg U_s=\mathrm{Const}$ drives the sequence of quantum phase transitions (QPT) where (pseudo)spin state on the lattice strongly change. There is a kind of frustration when $U_0\sim |U_s|$: the interactions drive the system to different phases when one interaction dominates the other one. While competition between the interactions leads to QPTs and induces new phases.

Vector cold atoms on optical lattices is the realisation of the system where $U_s$ is tunable. The origin of $U_s$-interaction in cold atom system can have different nature. For vector bosons it is typically related to difference in atom scattering lengthes for scattering with zero and nonzero total spin~\cite{Imambekov03}.  The magnitude and sign of atomic scattering length can be tuned, for example, by means of Feshbach resonance~\cite{George14_RoMP}. This manipulation of  (pseudo)spin-dependent scattering lengthes has become the area of active research~\cite{Galitski2014PRL,Meera2015PRL}.

We start our investigation with spin-$1$ bosons on the optical lattice governed by Bose-Hubbard model with intersite hopping energy $t$ and two types of interactions: $U_0 n_i(n_i-1)/2$ and $U_s\mathbf S_i^2/2$, where $n_i$ and $\mathbf S_i$ are particle density and spin at site $i$. Then in the Mott insulating state Bose-Hubbard model can be reduced to the effective lattice spin Hamiltonian using hopping $t\ll U_0,U_s$ as the small parameter~\cite{Imambekov03}:
\begin{gather}\label{eqH}
 H=\sum_{\langle i,j\rangle}(J_0+J_1 \mathbf S_i\cdot\mathbf S_j+J_2(\mathbf S_i\cdot\mathbf S_j)^2).
\end{gather}
Explicit expressions for for $J_{0,1,2}$ through hopping amplitudes $t$ and interactions $U_0$ and $U_s$ are well known~\cite{Imambekov03}. However physical properties of this system has been investigated only in the limit $U_s\ll U_0$.

We take $E_u= 2t^2/U_0$ as the energy unit. In general, exchange integrals behave with $\lambda=U_s/U_0$ as follows (at average atom filling $n=1$ of each lattice site): $J_1=2/(1+\lambda)$ and $J_2=2/(1-2\lambda)(1+\lambda)$. The divergencies show inaccuracy of $J_{1,2}$ at close proximity to the poles where more refined calculation (higher orders over $t/U_0$ and $t/U_s$) is necessary. However QPTs are sensitive mostly to the sign change of $J_{1,2}$ that we establish accurate enough.
\begin{figure}[t]
  \centering
  \includegraphics[width=\columnwidth]{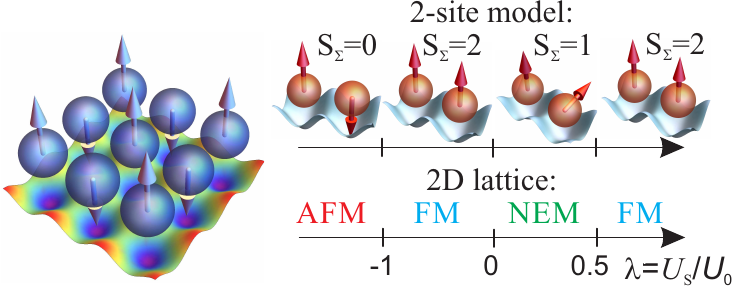}\\
  \caption{(Color online) Phase transitions induced by spin channel interaction $U_s$ in the optical trap consisting of two sites and in a two-dimensional lattice.  \label{Fig1}}
\end{figure}
\begin{figure}[t]
  \centering
  \includegraphics[width=0.9\columnwidth]{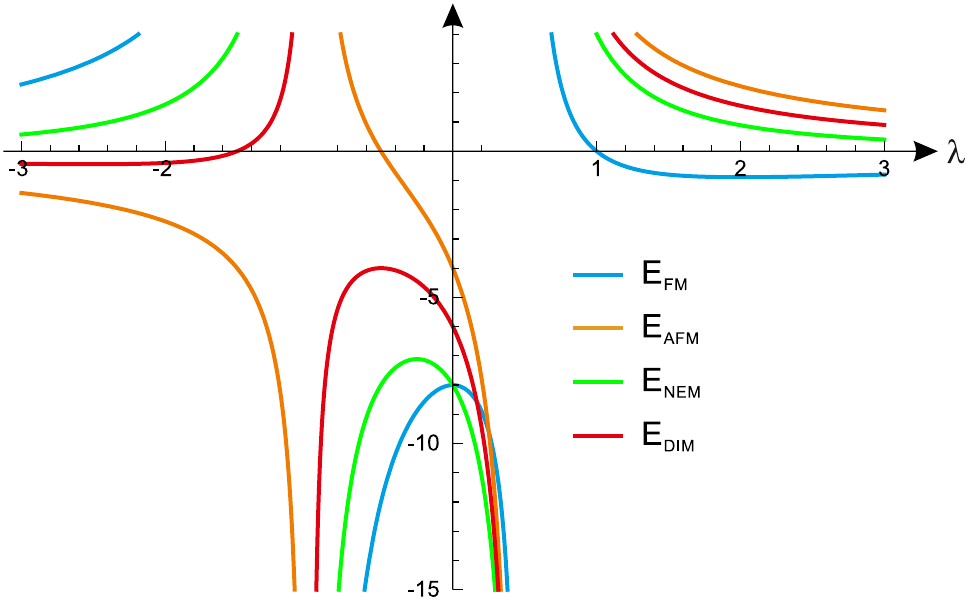}\\
  \caption{(Color online) Mean field energies of lattice spin Hamiltonian~\eqref{eqH} for square lattice (one atom per site).}\label{figDemler}
\end{figure}

To get an insight into the  phase diagram we first take the simple model, with only two sites, see Fig.~\ref{Fig1}. In the Mott phase quantum state of the system can be characterised by the total spin $\Ss=0,1,2$. It is well known how energies of  the states with $\Ss=0,1,2$ depend on the  triplet interaction amplitude $U_s$~\cite{Imambekov03}. In Fig.~\ref{Fig1} we show how the ground state  depends on $\lambda$ at the atom filling  $n=1$ (one atom per well): there are many QPTs changing the magnetic structure of the system. The phase diagram of vector bosons on the lattice at the mean field level  behaves very similarly: see lower diagram in Fig.~\ref{Fig1}.

On optical lattices spinor bosons in Mott insulator regime can form several distinct phases, which differ in their spin correlations: ferromagnetic, antiferromagnetic, nematic, dimer~\cite{Imambekov03}. Their energies are in the mean-field approximation~\cite{Supplementary}:
\begin{gather}\label{E1}
\EFM =\frac{\nu}{2}(J_0+ J_1+ J_2),
\,\,
\EAFM= \frac{\nu}{2}(J_0- J_1+ 2J_2),
\\\notag
\ENEM=\frac{\nu}{2}(J_0+ 2J_2),
\quad
\EDIM=\frac{\nu}{2} {\,} J_0 - J_1+ \frac23 J_2 (\nu+ 2),
\end{gather}
where $\nu= 2D$ is the number of nearest neighbors for $D$-dimensional cubic lattice.

Mean-field phase diagram of cold atoms in two-dimensional ($D=2$) lattice for the average lattice site filling $n=1$ is sketched schematically in Fig.~\ref{Fig1} using the graph~Fig.~\ref{figDemler} of mean field energies~\eqref{E1}. Exchange constants nonlinearly depend on $U_s$, so the quantum phase transitions take place not only at $U_s$ where $J_{1,2}$ change sign but like in the problem with two sites there is QPT at $\lambda=0$ induced by $U_s$ sign change. But there is no QPT at $\lambda=-1/4$ where $J_2$ changes its sign.

Similarly, at the filling $n=3$ (three atoms on average per lattice site), QPT between ferromagnetic and nematic states takes place at $\lambda=-1/4$ rather than $\lambda=0$.

\textit{As the second realisation} we consider Mott insulating state of fermions with spin-$1/2$ on the lattice, where fermion on each lattice site has in addition to spin  another  degree of freedom, the ``orbital'' one, described by the quantum numbers $\alpha=1,2$. This orbital degree of freedom in electron strongly correlated systems ($d$-electron compounds) corresponds to different choice of electron orbitals on each site~\cite{Kugel1982UFN,Yamashita1998PRB}. For cold atoms on the optical lattice this case also applies~\cite{li2013Nature}, but there are other realisations, e.g.,  when atom has a dipole moment (then $\alpha$ is its projection), or when the lattice site consists of two wells (then $\alpha$ labels atom position in the subwells)~\cite{Bruder2011PRA,Xu15_PRA,Chtchelkatchev2014PRA}. In most of these realisations the interaction part of Hubbard-like Hamiltonian describing this fermion system can be divided into two parts. The first term has a trivial structure in the well index space, and describes the Coulomb repulsion of fermions at one node: $\frac12U_0\sum_{i,\sigma,\sigma',\alpha,\alpha'} n_{i\alpha\sigma} n_{i,\alpha',\sigma'}(1-\delta_{\alpha\alpha'}\delta_{\sigma\sigma'})$. The second term usually  can be expressed like the Hund correlation energy~\cite{Kugel1982UFN}, $-U_s\sum_{i,\sigma,\sigma'}c_{i,1,\sigma}^\dag c_{i, 1,\sigma'} c_{i,2,\sigma'}^\dag c_{i,2,\sigma}$.

\begin{figure}[t]
  \centering
  \includegraphics[width=\columnwidth]{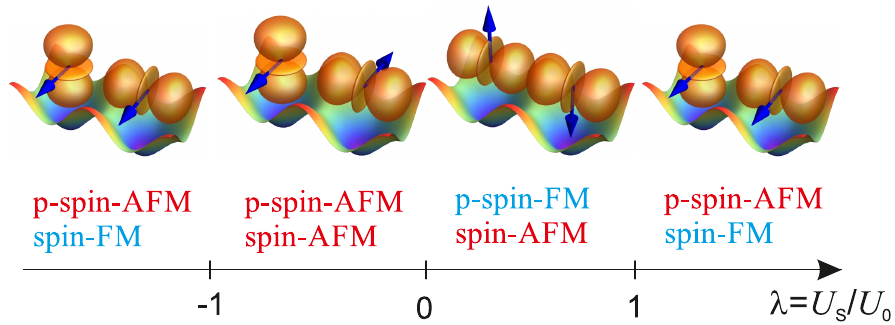}\\
  \caption{(Color online) Quantum phase transitions induced by interaction $U_s$ in the optical lattice with vector atoms having orbital degree of freedom. The average atom filling $n=1$ for each site. Pictures with atom p-orbitals  schematically show atoms occupying the nearest sites and arrows show spin. }\label{figKugelHomsky1}
\end{figure}
\begin{figure}[t]
  \centering
  \includegraphics[width=\columnwidth]{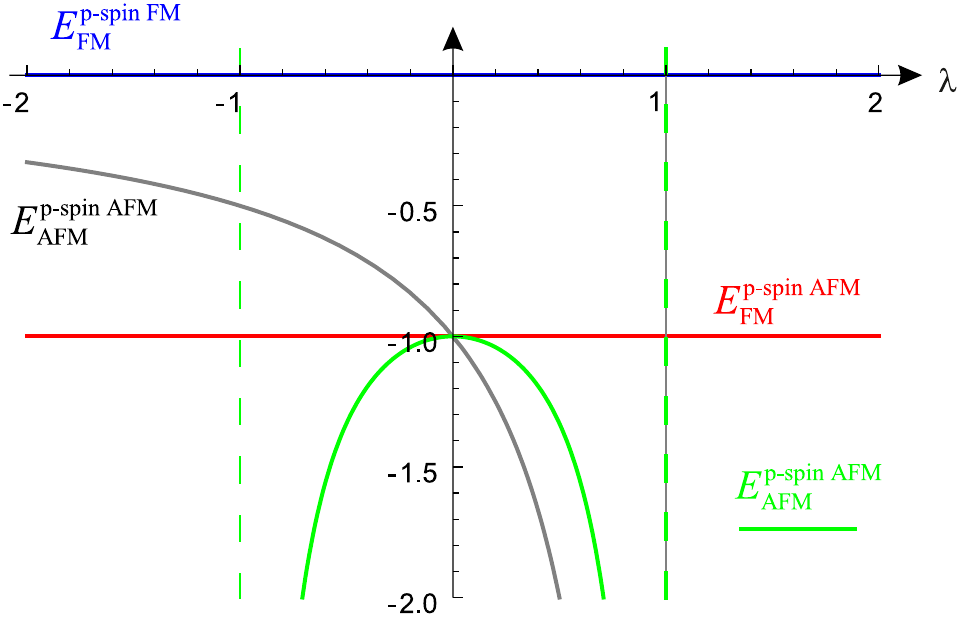}\\
  \caption{(Color online) Mean field energies of the Kugel-Khomskii Hamiltonian $\HKH$. Energy unit is $E_u= 2t^2/U_0$.}\label{figKugelHomsky2}
\end{figure}
The standard perturbative procedure over hopping amplitudes reduces the Hubbard model for fermion atoms to  the Kugel-Khomskii effective  Hamiltonian~\cite{kugel1973JETP, Kugel1982UFN,Yamashita1998PRB}: $\HKH=J_0\sum_{\langle ij\rangle}\{(\frac14 +\textbf{S}_i\cdot\textbf{S}_j)[J_1+J_2\bm{{\mathcal T}}_i\cdot\bm{\mathcal T}_j+J_3{\mathcal T}_i^z{\mathcal T}_j^z]+J_4\bm{{\mathcal T}}_i\cdot\bm{\mathcal T}_j+J_5{\mathcal T}_i^z{\mathcal T}_j^z\}$, where $\mathbf S_i$ is the fermion spin on the lattice site $i$ and $\bm{\mathcal T}_i$ is the pseudospin-$1/2$ operator describing ``orbital'' degree of freedom~\cite{Yamashita1998PRB}:
\begin{gather}\label{eqSc}
  S^a_i= c^\dag_{i\alpha \sigma} s^a_{\sigma\sigma'}c_{i\alpha \sigma'},
  \qquad
  {\mathcal T}^a_i= c^\dag_{i\alpha \sigma} \tau^a_{\alpha\beta} c_{i\beta \sigma},
\end{gather}
where $c_{i\alpha \sigma}$ is fermion annihilation operator, $s_{\sigma\sigma'}=\frac12\sigma^a_{\sigma\sigma'}$ is spin-$1/2$ operator  and $\sigma^a_{\sigma\sigma'}$ is the Pauli matrix. Similarly, $\tau^a_{\alpha\beta}=\frac12\sigma^a_{\alpha\beta}$. Doing the reduction we took the lattice with square (cubic) symmetry and average fermion density at a lattice node $\langle n_i \rangle=1$. Exchange integrals $J_a$, $a=0,\ldots,5$ of Kugel-Khomskii Hamiltonian have been found in all orders over $U_s/U_0$ in Ref.~\cite{kugel1973JETP} however the  properties  of its ground  state  were not investigated until now for  arbitrary  $\lambda=U_s/U_0$ (except $|\lambda|\ll 1$). The same applies for all anisotropic generalisations of this Hamiltonian. Looking at the explicit expressions for exchange integrals $J_a$ written in Eq.~6 of  Ref.~\cite{kugel1973JETP} and in Ref.~\cite{Yamashita1998PRB} we see that $J_0 =4 t^2/U_0$, $J_1=(1-\lambda^2-\lambda)/[2(1-\lambda^2)]$, $J_2=J_4/(2\lambda)=1/(1-\lambda^2)$ and $J_3=-J_5/2=2\lambda/(1+\lambda)$. So spin-spin and isospin-isospin exchange interactions change sign when $|U_s|\sim U_0$. This behaviour is the signature of a (quantum) phase transition where spin and isospin structures of  Mott insulating phase switch between ``ferro'' to ``antiferromagnetic'' phases, see Fig.~\ref{figKugelHomsky1}.

Again, there is  a QPT at $\lambda=0$ induced by $U_s$ sign change. It follows from the investigation of the mean field ground state energy of the Kugel-Khomskii Hamiltonian. Let’s consider first ferromagnetic (FM) pseudospin arrangement. Then $\bm{{\mathcal T}}_i\cdot\bm{\mathcal T}_j={\mathcal T}_i^z{\mathcal T}_j^z = 1/4$ and $\HKH$  reduces to $H=J_0\sum_{\langle ij\rangle}\{\textbf{S}_i\cdot\textbf{S}_j-1/4\}$. So the energies of pseudospin (``p-spin'') ferromagnetic and p-spin antiferromagnetic states are $E_{\mathrm{\scriptscriptstyle{FM}}}^{\rm p-spin\, \mathrm{\scriptscriptstyle{FM}}} = 0$, and $E_{\mathrm{\scriptscriptstyle{FM}}}^{\rm p-spin\, \mathrm{\scriptscriptstyle{AFM}}} = -\nu E_u/2$, where, like above, $E_u= 2t^2/U_0$ and  $\nu= 2D$. Let’s consider now antiferromagnetic (AFM) pseudospin arrangement. Then  $\bm{{\mathcal T}}_i\cdot\bm{\mathcal T}_j={\mathcal T}_i^z{\mathcal T}_j^z = -1/4$ and $\HKH$ reduces to  $H=-J_0\sum_{\langle ij\rangle}\{J_4\textbf{S}_i\cdot\textbf{S}_j+\frac{J_2+J_4}4\}$. So the energies of p-spin ferromagnetic and p-spin antiferromagnetic states are $E_{\mathrm{\scriptscriptstyle{AFM}}}^{\rm p-spin\, \mathrm{\scriptscriptstyle{AFM}}} =-\frac\nu 2E_u\frac1{1-\lambda}$, and $E_{\mathrm{\scriptscriptstyle{AFM}}}^{\rm p-spin\, \mathrm{\scriptscriptstyle{AFM}}} = -\frac\nu 2E_u\frac1{1-\lambda^2}$. The ground state energy corresponds to minimum of these four  energy states. We find it in Fig.~\ref{figKugelHomsky2} and sketch the phase diagram in Fig.~\ref{figKugelHomsky1}.

\textit{The third realisation} we investigate here is strongly interacting two-species bosons with spin $S = 1$ on optical lattice. In the Mott phase we observe the transition where atoms completely regroup: atom bunching into domains with single atom sort changes to the alternating arrangement of atoms. This transition is accompanied with the magnetic (spin) phase transition from ferromagnetic to antiferromagnetic state.

\begin{figure*}[t]
  \centering
  \includegraphics[width=\textwidth]{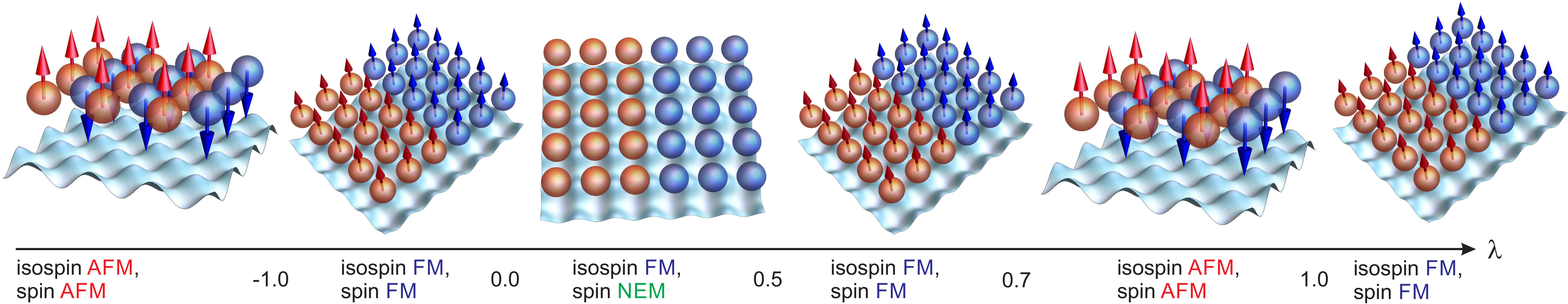}\\
  \caption{(Color online) A sketch of the phase diagram for $S=1$ bosons: QPTs are driven by spin channel interaction. At $\lambda= 0.5$ there is a transition from the phase with zero average magnetisation to spin-ferromagnetic phase. At larger value of the interaction, $\lambda=\frac{1}{10} \left(\sqrt{5}+5\right)\approx 0.7$ atoms reorder to the antiferromagnetic isospin state, with alternating arrangement of atoms, and the spin state again prefers zero average magnetisation. At higher strength of the interaction the system undergoes  the instability where atoms again regroup to the domain structure with ferromagnetic spin ordering in each domain. }\label{fig:lattices}
\end{figure*}
\begin{figure}[t]
  \centering
  \includegraphics[width=\columnwidth]{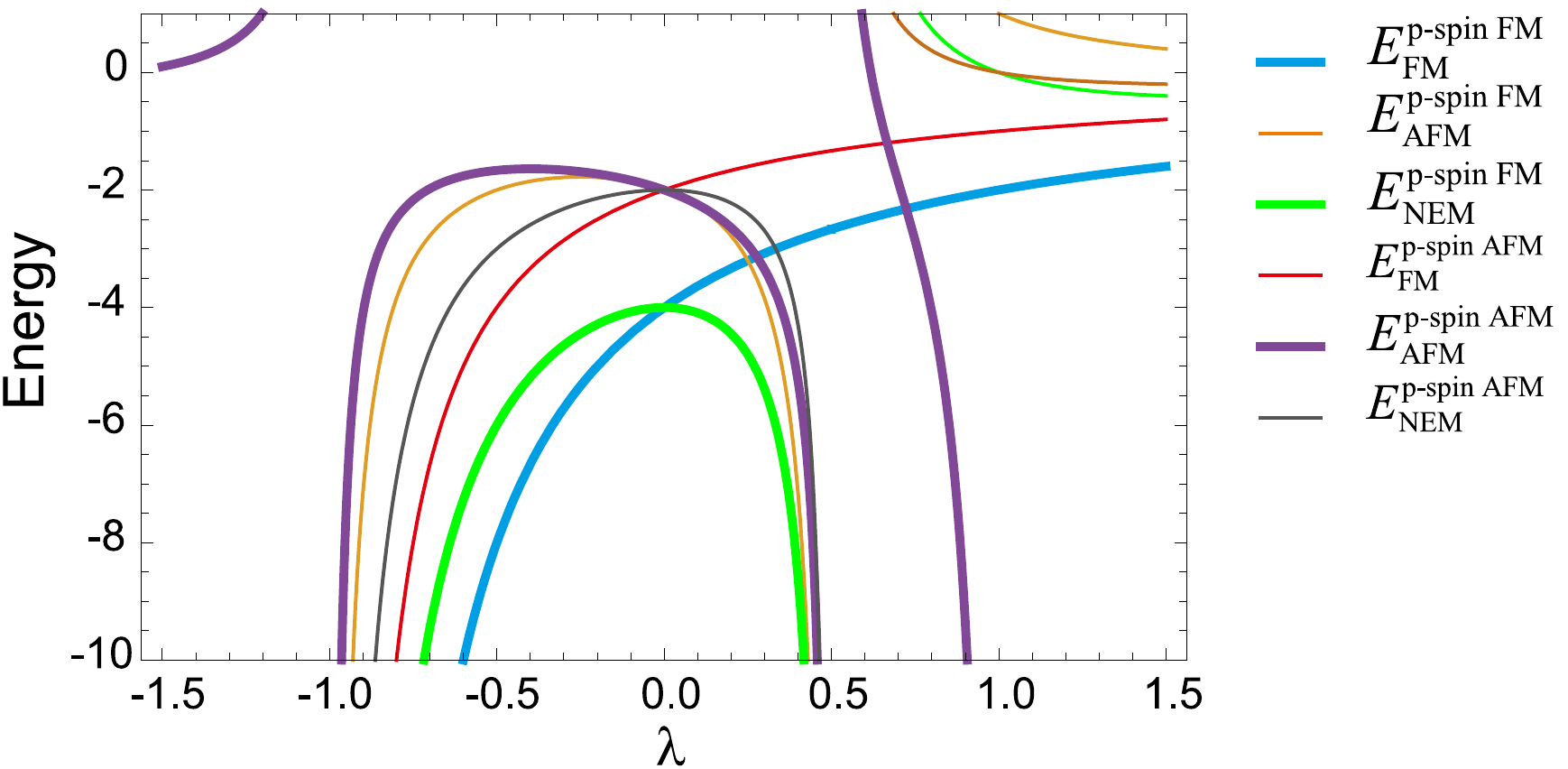}\\
  \caption{(Color online) Mean-field energies of $H_{\eff}$ corresponding to two sorts of $S=1$ bosons on the square optical lattice for the average site filling $\langle n_i \rangle= 1$ (the total number of bosons at site $i$). Energy is normalised by $E_u$.}\label{FigVectorBosons}
\end{figure}
For two types of boson atoms with $S= 1$ we introduce the creation operators $c^{\dagger}_{i \alpha s}$ for states localized on site $i$ and having spin components $s= \{ -1, 0, 1 \}$. Index $\alpha= 1, 2$ accounts for two types of bosons.   The interaction between bosons is given by two terms. The first one accounts for Hubbard repulsion between a pair of boson atoms residing at the same site~\cite{dutta2014RMP}: $2h_{U_0}=2U_{12} n_{i,1} n_{i,2} + U_{11} n_{i,1} (n_{i,1}- 1)+ U_{22}n_{i,2} (n_{i,2}- 1)$.  Here $U_{11}$, $U_{22}$ and $U_{12}$ are three interaction parameters.  A spin-dependent interaction term, originating from the difference in scattering lengths for $S= 0$ and $S= 2$ total spin channels, is of the form (see, e.g., \cite{Imambekov03, Tsuchiya04}) $2h_{U_s}= U_s( \mathbf S_i^2- 2n_i)$. Here $n_i= n_{i,1}+ n_{i,2}$ is the boson number at site $i$. The hopping term: $H_t= \sum_{\langle i,j \rangle, \alpha} t_{\alpha} \left( c^{\dagger}_{i \alpha s} c_{j \alpha s}+ c^{\dagger}_{j \alpha s} c_{i \alpha s} \right)$, where summation is performed over all links $\langle i,j \rangle$ and $\alpha=1,2$.

In the second order perturbation theory over $ t_{\alpha}$ the effective Hamiltonian has the form $H_{\eff}= H_t (E_G- H_0)^{-1} H_t$. Here $E_G= 0$ is the ground state energy of the Hamiltonian $H_0= \sum_i(h_{U_0}+h_{U_s})$ for the average site filling $\langle n_i \rangle= 1$. We can define spin $\mathbf S_i$ and pseudospin $\bm{\mathcal T}_i$ in a similar way like it has been done in Eq.~\eqref{eqSc} (but then $s_{\sigma\sigma'}$ is spin-1 operator) and express $H_{\eff}$ through them. The general expression is rather involved and so we will present it elsewhere. To illustrate new physical results without loss of generality  we focus on the limiting cases. We assume that bosons do not strongly differ from each other: $U_{12}\simeq U_{11}\simeq U_{22}=U_0$ and $t_1\simeq t_2=t$. Then~\cite{Supplementary}
\begin{gather} \notag
H_{\eff}= \sum \limits_{\langle i,j \rangle} \Bigl\{  \left[ \epsilon +  J {\bf S}_i \cdot {\bf S}_j+ K  ({\bf S}_i \cdot {\bf S}_j)^2 \right] \left[\frac{1}{2}+ 2 {\mathcal T}^z_i {\mathcal T}^z_j\right]+
\\\notag
\left[ \epsilon' +  J' {\bf S}_i \cdot {\bf S}_j+ K' ({\bf S}_i \cdot {\bf S}_j)^2 \right] \left[ \frac14- 2 {\mathcal T}^z_i {\mathcal T}^z_j+ {\bm {\mathcal T}}_i \cdot {\bm {\mathcal T}}_j \right] \Bigr\}.
\end{gather}
Effective exchange integrals are $J'= 2 \lambda/(1- \lambda^2)$, $K'=-2\lambda^2/(1- \lambda^2)(1- 2\lambda)$,
$\epsilon=2 \lambda/(1+\lambda)(1- 2\lambda)$,  and $\epsilon'=-2 (1- \lambda^2- 2\lambda)/(1-\lambda^2)(1- 2\lambda)$, where as above, $\lambda= U_s/U_0$ and the unit of energy is $E_u= 2t^2/U_0$. Here  $J$ and $K$ coincide with  $J_1$ and $J_2$ defined after Eq.~\eqref{eqH}. Note that $H_{\eff}$ involves the projectors in the pseudospin subspace: $P= \frac{1}{2}+ 2 {\mathcal T}^z_i {\mathcal T}^z_j$ and  $P'=\frac14- 2 {\mathcal T}^z_i {\mathcal T}^z_j+ {\bm {\mathcal T}}_i \cdot {\bm {\mathcal T}}_j $.

Below we discuss possible spin and pseudospin arrangements in the ground state of $H_{\eff}$. There are two basic types of pseudospin arrangements. The first one is realized when the lattice is filled with one type of bosons or when atoms of one sort bunch into domains on the lattice. Then we have ferromagnetic pseudospin wave function $|+ + \dots \rangle$ or $|- - \dots \rangle$.

The second type of pseudospin arrangement is realized when two types of bosons are alternating  on the neighboring sites. The pseudospin wave function is of the antiferromagnetic type $|+ - + - \dots \rangle$.

For both cases the effective Hamiltonian in spin subspace has the form like the Hamiltonian in Eq.~\eqref{eqH} but the corresponding parameters for the FM ($l= 1$) and AFM ($l= 2$) pseudospin arrangements are
\begin{gather}\notag
J_1^{(l=1)}= -E_u \frac{1}{1+ \lambda}, \quad J_2^{(l=1)}= -E_u \frac{1}{(1+ \lambda)(1- 2\lambda)},
 \\\label{P_1}
J_0^{(l=1)}= E_u \frac{2 \lambda}{(1+ \lambda)(1- 2\lambda)};
\\\notag
J_1^{(l=2)}= E_u \frac{\lambda}{1- \lambda^2}, \quad J_2^{(l=2)}= -E_u \frac{\lambda^2}{(1- \lambda^2)(1- 2\lambda)},
\\\label{P_2}
J_0^{(l=2)}= -E_u \frac{1- \lambda^2- 2 \lambda}{(1-\lambda^2)(1- 2\lambda)}.
\end{gather}

Explicit analytical expressions for the mean-field ground state energies of the effective Hamiltonian are  the same as given in Eq.~\eqref{E1}, but with exchange integrals~\eqref{P_1}-\eqref{P_2}.  As follows, for two-dimensional optical lattice ($D=2$), there is quantum phase transition at $\lambda=0$ when ferromagnetic spin state switches to nematic spin arrangement while the isopin state (atom sorts distribution over the lattice) does not change.

When $0 < \lambda < 0.5$, the lowest energy state corresponds to ``ferromagnetic'' isospin state: atoms of one sort cluster into domains, and zero magnetisation NEM spin state. When we cross the point $\lambda= 0.5$, and go higher over $\lambda$, we see that the ground state of the system is still ``ferromagnetic'' over isospin, but now also it is ferromagnetic over the real spin, see Fig.~\ref{fig:lattices}. So at $\lambda= 0.5$ atom spin spontaneously polarise on one direction.

At $\lambda=\frac{1}{10} \left(\sqrt{5}+5\right)\approx 0.7$ (and at $\lambda=-1$) atoms distribution over the lattice changes from alternating to clustering into domains. At the same time this transition is accompanied by the change of magnetic state.

To conclude, we consider quantum strongly correlated systems where frustration is induced by the competing interactions: they drive the system to different phases when one interaction dominates the other one, but their competition, in addition to localisation effects, leads to the origin of new phases and  cascade of quantum phase transitions.  Unlike condensed matter systems this situation is natural for cold atoms on optical lattices: we suggest three realisations. Our results will not only give the growing experimental interests in ultra-cold atomic gases, but also bring new perspectives in searching new materials in condensed matter systems.

N.C. acknowledges A. Petkovic for stimulating discussions at the initial stage of this work,  Laboratoire de Physique Théorique, Toulouse, where this work was initiated, for the hospitality and CNRS. This work was supported by the NSFC under grants Nos. 11434015, 61227902, SPRPCAS under grants No. XDB01020300 and partially funded by Russian Foundation for Scientific Research (grant No.16-02-00295), while numerical simulations were funded by Russian Scientific Foundation (grant No.14-12-01185). We thank Supercomputer Centers of Russian Academy of Sciences and National research Center Kurchatov Institute for access to URAL, JSCC and HPC supercomputer clusters.

\bibliography{refs}

\end{document}